\documentclass[%twocolumn,%showpacs,preprintnumbers,
amsmath,%amssymb,aps,
prd,nofootinbib,floatfix,12pt,%preprint,
]{revtex4}

\def\MSbar{\overline{\rm MS}}
\def\lnbar{\overline{\ln}}
\newcommand\beq{\begin{eqnarray}}
\newcommand\eeq{\end{eqnarray}}
\newcommand\Tbar{\overline{T}}

\allowdisplaybreaks
\usepackage{graphicx}% Include figure files
\usepackage{setspace}
\usepackage{bm}% bold math

\usepackage[titletoc,title]{appendix}
\begin{document}
\renewcommand{\theequation}{\arabic{section}.\arabic{equation}}
\renewcommand{\thefigure}{\arabic{section}.\arabic{figure}}
\renewcommand{\thetable}{\arabic{section}.\arabic{table}}

\title{\large 
Top-quark pole mass in the tadpole-free $\MSbar$ scheme}
\baselineskip=16pt 

\author{Stephen P.~Martin}
\affiliation{Department of Physics, Northern Illinois University, DeKalb IL 60115}

\begin{abstract}\normalsize\baselineskip=16pt 
The complex pole mass of the top quark is presented at full two-loop order 
in the Standard Model, augmenting 
the known four-loop QCD contributions.
The input parameters are the $\MSbar$  
Yukawa and gauge couplings, the Higgs 
self-coupling, and the Higgs vacuum expectation value (VEV). 
Here, the VEV is defined
as the minimum of the full effective potential in Landau gauge, 
so that tadpoles vanish. This is an alternative to earlier results that
instead minimize the tree-level potential, resulting in a VEV that is 
gauge-fixing independent but accompanied by negative
powers of the Higgs self-coupling in  
perturbative expansions. The effects of non-zero Goldstone 
boson mass are eliminated by resummation. I also study the renormalization scale 
dependence of the calculated pole mass. 
\end{abstract} 
\maketitle

\vspace{-0.45in}

\tableofcontents

\baselineskip=16pt

\vspace{-0.15in}

%%%%%%%%%%%%%%%%%%%%%%%%%%%%%%%%%%%%%%%%%%%%%%%%%%%%%%%%%%%%%%%%
\section{Introduction\label{sec:intro}}
\setcounter{equation}{0}
\setcounter{figure}{0}
\setcounter{table}{0}
\setcounter{footnote}{1}

The top-quark mass is one of the key parameters of the Standard Model of particle 
physics. It is important for precision electroweak fits, and for matching of the
Standard Model to ultraviolet physics, including both stability of the electroweak vacuum
and theories that attempt to address the hierarchy problem.

In this paper, I will consider the
relation between the $\MSbar$ Lagrangian quantities and the complex pole squared
mass, which formally is a physical observable 
\cite{Tarrach:1980up,Passera:1998uj,Kronfeld:1998di,Gambino:1999ai} and
does not depend on the choice of gauge fixing terms or on the 
renormalization scheme and the renormalization
scale  to all orders in perturbation theory, but
is subject to non-perturbative
renormalon ambiguities associated with the hadronization scale
\cite{Bigi:1994em,Beneke:1994sw}.
The relationship between the pole mass and the top-quark mass
as measured by hadron collider experiment collaborations is somewhat problematic
and is the subject of continuing investigations \cite{Mtopcollider}. 
In the approximation that
the width of the top-quark is neglected, the real part of the complex pole squared mass
coincides with the on-shell squared mass. 
There exist several other useful definitions of the top-quark mass, 
depending on the precise relation to experimental quantities. These include the 
the potential-subtracted  mass \cite{Beneke:1998rk}, 
the 1S mass \cite{oneSmassbottom,oneSmasstop}, and the running $\MSbar$ mass. 

In this paper, I consider the relation between the top-quark pole mass
and the the $\MSbar$ Lagrangian quantities. Many previous works have 
contributed to this subject. First, the pure QCD contributions have been given at 1-loop 
order \cite{Tarrach:1980up}, 2-loop order \cite{Gray:1990yh} (confirmed 
in \cite{Avdeev:1997sz,Fleischer:1998dw}), 
3-loop order \cite{Melnikov:2000qh} (with 
previous approximate results in \cite{Chetyrkin:1997wm,Chetyrkin:1999qi}, and a useful summary of formulas in \cite{RunDec}), and 4-loop 
order \cite{Marquard:2015qpa}. Besides these pure QCD contributions,
the full 1-loop contributions to the pole mass have been
given in ref.~\cite{Bohm:1986rj} (see also ref.~\cite{Hempfling:1994ar} and eq.~(B.5) of ref.~\cite{Jegerlehner:2002em}).
The 2-loop mixed QCD contributions were found in \cite{Jegerlehner:2003py},
and confirmed in ref.~\cite{Eiras:2005yt}.
The full 2-loop contributions have been studied 
in the gaugeless limit (with the electroweak vector boson masses neglected 
compared to the top-quark mass) in
refs.~\cite{Faisst:2003px,Jegerlehner:2003sp,Faisst:2004gn,Martin:2005ch,Kniehl:2014yia}. 
Most recently the full 2-loop results were given in ref.~\cite{Kniehl:2015nwa}.

The purpose of the present paper is to give an alternative calculation of the full 2-loop
contributions to the top-quark pole mass, using a different organization of 
perturbation theory than in the above references. Note that
the definition of the running $\MSbar$ top-quark mass is not 
unique for a given renormalization scale $Q$, 
because the mass is proportional to the Higgs VEV, which can be 
defined in more than one way. One way, called the ``tree-level VEV" 
here, is 
\beq
v_{\rm tree} = \sqrt{-m^2/\lambda},
\eeq
where $\lambda$ and $m^2$ are the Higgs self-coupling and squared mass 
parameter in the $\MSbar$ scheme, normalized so that the tree-level 
$\MSbar$ renormalized potential for the canonically normalized
complex Higgs doublet field $\Phi$ is
\beq
V(\Phi, \Phi^\dagger) &=& m^2 \Phi^\dagger \Phi + \lambda (\Phi^\dagger \Phi)^2 .
\eeq
An advantage of $v_{\rm tree}$, emphasized for example in 
refs.~\cite{Jegerlehner:2012kn,Kniehl:2015nwa}, 
is that it and the corresponding 
tree-level mass $m_t = y_t v_{\rm tree}/\sqrt{2}$ are manifestly independent of the 
gauge-fixing procedure, due to the way that they are defined 
in terms of the $\MSbar$ Lagrangian
parameters. A disadvantage is that, although there are no tree-level tadpoles,
there are tadpole loop diagrams involving the Higgs field, which have to be 
included in any calculation based on $v_{\rm tree}$ or $m_t$. 
As a consequence, the perturbative loop expansion parameters include 
\beq
N_c y_t^4/(16 \pi^2 \lambda)
\eeq
rather than the usual $N_c y_t^2/16 \pi^2$. The presence of powers of 
$\lambda \approx 0.126$ in the 
denominators of perturbative expansions is due to the tadpoles, and
is indicative of the fact that the tree-level VEV is 
not a very good approximation for the true vacuum state
of the theory after including loop corrections. For example,
in ref.~\cite{Jegerlehner:2012kn}, it was noted that when using $v_{\rm tree}$,
the 1-loop non-QCD correction is surprisingly huge, almost canceling the
1-loop QCD effect, due to 
the $1/\lambda$ tadpole effects.

In this paper, I follow the alternative scheme of defining 
the running $\MSbar$ squared masses of the top quark, bottom quark,
electroweak vector bosons,
and the Higgs scalar boson by
\beq
t &=& y_t^2 v^2/2,
\label{eq:deft}
\\
b &=& y_b^2 v^2/2,
\\
W &=& g^2 v^2/4,
\\
Z &=& (g^2 + g^{\prime 2}) v^2/4,
\\
h &=& 2 \lambda v^2
\label{eq:defh}
\eeq
where the the VEV $v$ of the Higgs field is defined to be the minimum of the 
full effective potential in Landau gauge. As a benefit of this 
definition, the sum of all Higgs tadpole graphs, including the tree-level 
Higgs tadpole, vanishes identically. The price to be paid for this is that the 
VEV $v$, and therefore also $t$ and the other tree-level masses, 
depend on the gauge-fixing method. 
Therefore, calculations based on $v$ are restricted to Landau gauge in the electroweak
sector (or any other gauge-fixing choice; Landau gauge is chosen 
only because the effective potential is simple).
Although one therefore apparently loses the check of requiring independence of 
the gauge-fixing parameters, the checks obtained from the 
cancellation of the unphysical Landau gauge Goldstone boson degrees of freedom from 
the complex pole squared mass (and other observables) 
are just as powerful. A benefit of this 
definition of the VEV is that $v$ is in some sense a more faithful description
of the true vacuum state. Negative powers of 
$\lambda$ are absent in perturbative expansions of pole masses and other 
physical quantities. Indeed, this provides another useful check. 

As a practical matter, the Standard Model effective potential is now known at 
full 2-loop order \cite{FJJ}, together with the 3-loop contributions in the 
approximation that QCD and top-Yukawa couplings are large compared to 
all other couplings \cite{Martin:2013gka}, and the 4-loop contributions 
at leading order in QCD \cite{Martin:2015eia}, and with 
resummation of 
the Goldstone boson contributions 
\cite{Martin:2014bca,Elias-Miro:2014pca} (see also \cite{Pilaftsis:2015bbs})
to avoid spurious infrared 
singularities and imaginary parts. As a consequence, one can write a   
loop expansion for the relationship between the two VEVs, showing the
tadpole contributions explicitly,
\beq
v_{\rm tree}^2 = v^2 + \frac{1}{\lambda} \sum_{\ell = 1}^\infty 
\frac{1}{(16\pi^2)^\ell}
\widehat \Delta_\ell,
\label{eq:translateVEVs}
\eeq
where $\widehat \Delta_1$ and $\widehat \Delta_2$ are known exactly, and 
$\widehat \Delta_3$ is known in the approximation that the QCD coupling and 
top Yukawa coupling are much larger than the other couplings. They are given in
eqs.~(4.19)-(4.21) of ref.~\cite{Martin:2014bca}. Also, $\widehat \Delta_4$ 
is known only at leading order in QCD; this is given in eq.~(5.5) of 
ref.~\cite{Martin:2015eia}.

The methods and results of the present paper are designed to be compatible 
with similar full 2-loop calculations of the complex pole squared masses 
of the Higgs scalar (with leading 3-loop contributions) in 
ref.~\cite{Martin:2014cxa}, the $W$ boson in ref.~\cite{Martin:2015lxa}, 
and the $Z$ boson in ref.~\cite{Martin:2015rea}. All of these use the 
same VEV definition $v$, as an alternative to similar results that are 
expressed in terms of $v_{\rm tree}$, or parameterized in terms of other
quantities such as the Fermi constant $G_F$. 
For important previous results on the electroweak vector boson masses and the Higgs mass
in the Standard Model using other schemes, see 
\cite{Sirlin:1980nh}-\cite{Kniehl:2016enc}.
  
The input parameters here are taken to be the $\MSbar$ top-quark and bottom-quark 
Yukawa couplings, the 
$SU(3)_c$, $SU(2)_L$
and $U(1)_Y$ gauge couplings, the Higgs self-coupling, and the VEV as discussed above:
\beq
y_t,\> y_b,\>  g_3,\>  g,\>  g',\>  \lambda,\>  v,
\eeq
all at a specified renormalization scale $Q$. 
In principle, the result also depends on the lighter quark and lepton Yukawa couplings 
(or masses), but their contributions are very small, 
with the largest contribution coming from the 2-loop 
QCD contribution of the bottom quark mass, as noted below. 
The effects of CKM mixing on the top-quark pole mass calculation are also
negligible. The $W$, $Z$, and $h$ 
physical masses, as well as quantities such as 
$G_F$ and $\sin^2\theta_W$, are all regarded as output 
quantities in this pure $\MSbar$ scheme adopted here.

\begin{figure}[]
\begin{center}
\includegraphics[width=0.98\linewidth,angle=0]{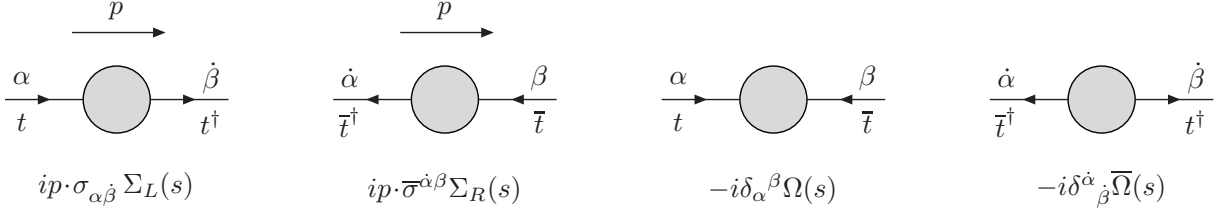}
\begin{minipage}[]{0.98\linewidth}
\caption{\label{fig:selfenergies}
Fermion one-particle-irreducible self-energy functions
$\Sigma_L(s)$, $\Sigma_R(s)$, $\Omega(s)$, and $\overline \Omega(s)$, using 2-component notation following ref.~\cite{Dreiner:2008tw}. Here the external momentum invariant
is $s = -p^2$, using a metric
signature ($-$,$+$,$+$,$+$). The 2-component fields $t$ and $\overline t$
are the left-handed $SU(2)_L$ doublet quark and singlet antiquark, 
respectively, and $t$, $t^\dagger$, $\overline t$, 
and $\overline t^\dagger$ are labeled as ingoing.}
\end{minipage}
\end{center}
\end{figure}

In the following, the complex pole squared mass\footnote{The pole 
squared mass is sometimes parameterized instead as
$s_{\rm pole} = (M_t' - i \Gamma'_t/2)^2$,
but $M_t'$ exceeds $M_t$ by less than 2 MeV, which is negligible
compared to both experimental and theoretical uncertainties.\label{foot:Mtp}}
is denoted by
\beq
s_{\rm pole} = M_t^2 - i \Gamma_t M_t .
\eeq 
Methods for calculating the complex pole mass at higher orders in 
perturbation theory from the fermion self-energy components are 
well-known, and given in various ways in several of the references mentioned above.
In this paper, I use the 2-component fermion notation of 
ref.~\cite{Dreiner:2008tw}, and followed the procedure outlined in 
\cite{Martin:2005ch}. Defining the 2-component fermion self-energy functions as in
figure \ref{fig:selfenergies}, the complex pole mass is the solution of
%%%%%%%%%%%%%%%%%%%%%%%%%%%%%%%%%%%%%%%%%%%%%%%%%%%%%%%%%%%%%%%%%%%%%
%%%%%%%%%%%%%%%%%%%%%%%%%%%%%%%%%%%%%%%%%%%%%%%%%%%%%%%%%%%%%%%%%%%%%
\beq
0 &=& {\rm Det}\left (s_{\rm pole} - 
[1 - \Sigma_L(s_{\rm pole})]^{-1} 
[m + \overline \Omega(s_{\rm pole})] 
[1 - \Sigma_R(s_{\rm pole})]^{-1} 
[m + \Omega(s_{\rm pole})] \right ),\phantom{xxx}
\eeq
where $m$ is the tree-level quark mass.
In the case of the top quark with the approximation of this paper that the
CKM mixing is absent, the self-energy functions are numbers, not matrices in flavor space, and the absence of complex couplings implies that $\overline \Omega = \Omega$.
The loop expansion of the self-energy functions is:
\beq
\Sigma_L &=& \frac{1}{16 \pi^2} \Sigma_L^{(1)} +  \frac{1}{(16 \pi^2)^2} \Sigma_L^{(2)} +
\ldots
\\
\Sigma_R &=& \frac{1}{16 \pi^2} \Sigma_R^{(1)} +  \frac{1}{(16 \pi^2)^2} \Sigma_R^{(2)} +
\ldots
\\
\Omega &=& \frac{1}{16 \pi^2} \Omega^{(1)} +  \frac{1}{(16 \pi^2)^2} \Omega^{(2)} +
\ldots ,
\eeq
and it follows that the complex pole mass is the solution of
\beq
s_{\rm pole} &=& m^2 
+  \frac{1}{16 \pi^2}\Pi^{(1)} (s_{\rm pole}) 
+ \frac{1}{(16 \pi^2)^2}\Pi^{(2)} (s_{\rm pole}) + \ldots,
\eeq
where
\beq
\Pi^{(1)} (s) &=& m^2 \bigl [
\Sigma_L^{(1)} (s) + \Sigma_R^{(1)} (s) \bigr ] + 2 m \Omega^{(1)}(s) ,
\\
\Pi^{(2)} (s) &=& m^2 \bigl [
\Sigma_L^{(2)} (s) + \Sigma_R^{(2)} (s) +\Sigma_L^{(1)} (s) \Sigma_R^{(1)} (s) 
+ (\Sigma_L^{(1)}(s))^2 + (\Sigma_R^{(1)}(s))^2
\bigr ]
\nonumber \\ &&
+ 2 m \bigl [ \Omega^{(2)}(s) +  \Omega^{(1)}(s) (\Sigma_L^{(1)} (s) + \Sigma_R^{(1)} (s)) \bigr ]
+ \bigl [\Omega^{(1)}(s) \bigr ]^2 .
\eeq
Now, expanding $s_{\rm pole}$ about $m^2$, one obtains
\beq
s_{\rm pole} = m^2 +  \frac{1}{16 \pi^2}\Pi^{(1)} (m^2)
+ \frac{1}{(16 \pi^2)^2} \left [\Pi^{(2)} (m^2) 
+  \Pi^{(1)} (m^2) \>\Pi^{(1)\prime} (m^2)\right ]
+\ldots,
\eeq

The remaining part of the calculation is very similar to the strategy 
used in refs.~\cite{Martin:2014cxa,Martin:2015lxa,Martin:2015rea}, so 
the reader is referred to those papers for more details, and only a 
brief outline will be given here. The fermion self-energy functions are 
computed in terms of bare couplings and masses in $d= 4 - 2 \epsilon$ 
dimensions, and then the bare quantities are expanded in terms of 
renormalized $\MSbar$ quantities. This is more efficient than doing 
counterterm diagrams separately. The Tarasov algorithm 
\cite{Tarasov:1997kx} is then used to reduce the loop integrals to a set 
of basis functions, for which I use the conventions and notations of 
refs.~\cite{Martin:2003qz,TSIL}, with 1-loop basis integrals $A(x)\equiv 
x \lnbar(x) - x$ and $B(x,y)$ and 2-loop basis integrals $I(x,y,z)$, 
$S(x,y,z)$, $T(x,y,z)$, $\overline T(0,x,y)$, $U(x,y,z,u)$, and 
$M(x,y,z,u,v)$. Here $x,y,z\ldots$ are squared mass arguments, and there 
are also implicit arguments for the external momentum invariant $s$ and the 
renormalization scale $Q$, and
\beq
\lnbar(x) \equiv \ln(x/Q^2)
.
\eeq
The software package {\tt TSIL} \cite{TSIL} 
is used to evaluate the basis integral functions.
In some cases, the basis integrals can be evaluated analytically in terms of 
polylogarithms, which {\tt TSIL} does using results from 
refs.~\cite{Gray:1990yh,
Fleischer:1998dw,
Jegerlehner:2003py,
Martin:2003qz, 
TSIL,
Broadhurst:1987ei,
Scharf:1993ds,
Berends:1994ed,
Berends:1997vk,
Davydychev:1998si,
Fleischer:1999hp,
Martin:2003it}.
When this is not possible, {\tt TSIL} instead computes the basis 
integrals by Runge-Kutta integration of the differential equations in 
the external momentum invariant $s$ as found in 
ref.~\cite{Martin:2003qz}, using methods similar to those in 
refs.~\cite{Caffo:1998du}. The 1-loop self-energy functions of $s$ are 
then expanded around the tree-level squared mass $t$. Due to the 
presence of the massless gluon and photon, this expansion results in 
threshold logarithms $\lnbar(t - s)$, which cancel against 2-loop 
contributions, providing a useful check. Another important and 
non-trivial check is the cancellation of poles in $\epsilon$. 

\begin{figure}[]
\begin{center}
\includegraphics[width=0.90\linewidth,angle=0]{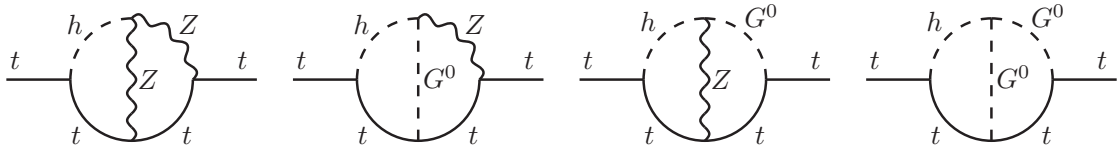}
\begin{minipage}[]{0.98\linewidth}
\caption{Feynman diagrams that individually involve the basis integrals
$M(h,Z,t,t,0)$ and $M(h,0,t,t,Z)$ and $M(h,0,t,t,0)$, 
coming from the unphysical
degrees of freedom from the $Z$ and the neutral Goldstone boson. 
In the total pole squared mass, the contributions proportional to each of
those basis integrals cancel.
\label{fig:example}}
\end{minipage}
\end{center}
\end{figure}

The resulting expression depends on the tree-level
Goldstone boson squared mass $G = 
m^2 + \lambda v^2$ and the Higgs boson squared mass $H = m^2 + 3 \lambda 
v^2$. Following the procedure in section IV of 
ref.~\cite{Martin:2014bca}, the Goldstone boson squared mass 
contributions are resummed, eliminating $G$ completely, and eliminating 
$H$ in favor of $h$ defined in eq.~(\ref{eq:defh}), so that the Higgs boson
squared mass parameter $m^2$ does not appear in the resulting expression.
Another useful check is the 
absence of singularities as $G \rightarrow 0$. Finally, in the remaining 
expressions, yet another useful check is provided by the cancellation 
between the Goldstone contributions and the unphysical components of the 
electroweak vector bosons. For one example, consider the Feynman 
diagrams shown in figure \ref{fig:example}. The contributions to $s_{\rm 
pole}$ from these individual diagrams from the neutral Goldstone boson 
$G^0$ and from the unphysical degrees of freedom of the $Z$ boson 
involve the basis integrals $M(h,Z,t,t,0)$ and $M(h,0,t,t,Z)$ and 
$M(h,0,t,t,0)$. However, in the sum, those unphysical contributions 
cancel, and these basis integrals do not appear at all in the result for 
the pole squared mass.
%%%%%%%%%%%%%%%%%%%%%%%%%%%%%%%%%%%%%%%%%%%%%%%%%%%%%%
%%%%%%%%%%%%%%%%%%%%%%%%%%%%%%%%%%%%%%%%%%%%%%%%%%%%%%
%%%%%%%%%%%%%%%%%%%%%%%%%%%%%%%%%%%%%%%%%%%%%%%%%%%%%%
This, along with many other similar cases, illustrates how gauge invariance 
provides non-trivial checks, despite the calculation being 
restricted to Landau gauge.

The rest of this paper is organized as follows. In section 
\ref{sec:Mtop}, results are presented for the full 2-loop complex pole 
mass of the top-quark, together with a review of the known 4-loop pure 
QCD results in a compatible form, using an expansion in which the 
external momentum invariant for the loop integrals is the tree-level 
squared mass $t$. In section \ref{sec:expandaroundpole}, the same 
results are presented after a re-expansion in which the external 
momentum invariant for the loop basis integrals is (the real part of) 
the top-quark pole squared mass. These two expansions differ by 
amounts that are formally of 5-loop order in pure QCD and 3-loop order 
for other terms. In section \ref{sec:num}, I compare these two 
expansions for a realistic set of numerical input parameters, which 
provides a test of the unavoidable arbitrariness associated with the 
truncation of perturbation theory. I also study the renormalization 
scale dependence of the approximation. Section \ref{sec:outlook} 
contains some concluding remarks. Some integral identities that were 
useful for the calculation are recorded in an Appendix.

%%%%%%%%%%%%%%%%%%%%%%%%%%%%%%%%%%%%%%%%%%%%%%%%%%%%%%%%%%%%%%%%
\section{Complex pole squared mass of the top-quark\label{sec:Mtop}}
\setcounter{equation}{0}
\setcounter{figure}{0}
\setcounter{table}{0}
\setcounter{footnote}{1}

In this section, the complex pole squared mass of the top-quark is 
written in the form:
\beq
s_{\rm pole} &=& t \,+\, 
\frac{1}{16\pi^2} \bigl [g_3^2 \delta^{(1)}_{\rm QCD} + \delta^{(1)}_{\rm non-QCD} \bigr ]
\,+\, 
\frac{1}{(16\pi^2)^2} \left [g_3^4 \delta^{(2)}_{\rm QCD} + g_3^2 \delta^{(2)}_{\rm mixed} + \delta^{(2)}_{\rm non-QCD}
\right ]
\nonumber \\ &&
\,+\, 
\frac{1}{(16\pi^2)^3} g_3^6 \delta^{(3)}_{\rm QCD} 
\,+\, 
\frac{1}{(16\pi^2)^4} g_3^8 \delta^{(4)}_{\rm QCD} + \ldots ,
\label{eq:tpolet}
\eeq
where $t = y_t^2 v^2/2$ is the tree-level squared mass, with $v$ taken to minimize the
full loop-corrected Landau gauge 
effective potential. In this section, all of the basis loop integrals
are implicitly taken to be evaluated with the external
squared momentum invariant set to $s=t$.
I begin by reviewing the known pure QCD results. At 1-loop and 2-loop order, one has
\cite{Tarrach:1980up,Gray:1990yh}:
\beq
\delta^{(1)}_{\rm QCD} &=& 
C_F t [8 - 6 \lnbar(t)] ,
\\
\delta^{(2)}_{\rm QCD} &=& C_F t \Bigl [
C_G \Bigl ( \frac{1111}{12} - \frac{8 \pi^2}{3} + 8 \pi^2 \ln(2) - 12 \zeta_3
- \frac{185}{3} \lnbar(t) + 11 \lnbar^2(t)
\Bigr )
\nonumber \\ &&
+
C_F \Bigl (-\frac{7}{4} + 10 \pi^2 - 16 \pi^2 \ln(2) + 
24 \zeta_3 - 15 \lnbar(t) + 18 \lnbar^2 (t) \Bigr )
\nonumber \\ &&
+ T_F \sum_{i=1}^{n_q} 
\Bigl \{
4 (1 - q_i/t) \bigl [T(q_i,q_i,t) + \lnbar(q_i)
[2 - \lnbar(t)] \bigr ]
\nonumber \\ &&
-4 (1 + q_i/t) U(t,0,q_i,q_i) 
- 4 (q_i/t) [2 \lnbar(q_i) + 1] + [4 \lnbar(t) + 1]/3
\Bigr \}
\Bigr ] ,
\eeq
where the QCD group theory quantities are
\beq
C_G = 3,\qquad C_F =4/3,\qquad T_F =1/2,\qquad n_q = 6,  
\eeq
and the $q_i$ are the quark squared masses. For $q_i = t$, one has (for $s=t$):
\beq
%T(t,t,t) &=& -\frac{1}{2} - \lnbar(t) + \frac{1}{2} \lnbar^2(t)
%,
%\nonumber \\ 
U(t,0,t,t) &=& \frac{11}{2} - \frac{2\pi^2}{3} - 3 \lnbar(t) + \frac{1}{2} \lnbar^2(t)
.
\eeq
When $q_i$ is a lighter quark squared mass, the integrals 
$T(q,q,t)$ and $U(t,0,q,q)$ for $s=t$ are known 
\cite{Gray:1990yh,Davydychev:1998si,Berends:1997vk,TSIL}
in terms of dilogarithms, 
but it is practical to expand them for small quark masses:
\beq
T(q,q,t) &=& -\lnbar(q) [2 - \lnbar(t)] - \frac{1}{2} - \frac{\pi^2}{3}
+ \lnbar(t) - \frac{1}{2} \lnbar^2(t) + {\cal O}(q/t),
\\
U(t,0,q,q) &=& \frac{11}{2} + \frac{\pi^2}{3} - 3 \lnbar(t) + \frac{1}{2}\lnbar^2(t)
- 2\pi^2 \sqrt{q/t}
 + {\cal O}(q/t) .
\eeq
The leading correction due to small non-zero quark masses thus is of order
$m_q/m_t$ and comes only from the 
$U(t,0,q,q)$ integral. Plugging in the group theory quantities, and keeping 
only the leading part in the light-quark mass expansion, one obtains:
\beq
\delta^{(2)}_{\rm QCD} &=& t \Bigl [
2309/9 + 16 \pi^2/9 + 32 \pi^2 \ln(2)/9 - 16 \zeta_3/3 - 204 \lnbar(t) + 60 \lnbar^2(t)
\nonumber \\ &&
+ \frac{16 \pi^2}{3} \sum_{q=b,c,s\ldots} m_q/ m_t
+ {\cal O}(m_q^2/m_t^2) \Bigr ].
\label{eq:deltatwoQCD}
\eeq
Here it is not clear whether it is best to use pole or running masses 
for $m_q$, since the resulting difference for $s_{\rm pole}$ between 
these choices is of the same parametric order as the presently unknown 
dependence of the 3-loop corrections on the lighter quark masses. 
However, even if one uses $\sum_q m_q = 7$ GeV, the 
net effect is to raise the top-quark pole mass by only about 14 MeV, 
which is small compared to both experimental and other theoretical 
uncertainties.

The 3-loop and 4-loop pure-QCD contributions can be written in the forms:
\beq
\delta^{(3)}_{\rm QCD} &=& t \Bigl [c_{3,0} + c_{3,1} \lnbar(t)
 + c_{3,2} \lnbar^2(t) + c_{3,3} \lnbar^3(t) \Bigr ],
\\
\delta^{(4)}_{\rm QCD} &=& t \Bigl [c_{4,0} + c_{4,1} \lnbar(t)
 + c_{4,2} \lnbar^2(t) + c_{4,3} \lnbar^3(t) + c_{4,4} \lnbar^4(t) \Bigr ] ,
\eeq
where the 3-loop results from \cite{Melnikov:2000qh} are
\beq
c_{3,0} &=& 551909/81 + 1589684 \pi^2/1215 + 
            700 \pi^4/81 - 42304 \pi^2 \ln(2)/81
\nonumber \\ &&
- 512 \pi^2 \ln^2(2)/27 
-320 \ln^4(2)/9 - (2560/3) {\rm Li}_4(1/2) - 13328 \zeta_3/9 
\nonumber \\ &&
- 11512 \pi^2 \zeta_3/27 
+ 31600 \zeta_5/27 ,
\\
c_{3,1} &=& -53696/9 - 352 \pi^2/9 - 704 \pi^2 \ln(2)/9 + 2272 \zeta_3/3 ,
\\
c_{3,2} &=& 2300 ,
\\
c_{3,3} &=& -440 ,
\eeq
while the 4-loop coefficients are  \cite{Marquard:2015qpa}
\beq
c_{4,0} &=& (4.91 \pm 0.11) \times 10^{5} ,
\label{eq:cfourzero}
\\
c_{4,1} &=&  -5110172/27 - 46066276 \pi^2/1215 - 26348 \pi^4/81 + 
1231424 \pi^2 \ln(2)/81
\nonumber \\ &&
+ 14848 \pi^2 \ln^2(2)/27 + 9280 \ln^4(2)/9 + (74240/3) {\rm Li}_4(1/2)  
\nonumber \\ &&
+ 1824176 \zeta_3/27 + 333848 \pi^2 \zeta_3/27 - 1103600 \zeta_5/27,
\\
c_{4,2} &=&  750374/9 + 5104 \pi^2/9 + 320 \pi^2 \ln(2)/9 + 3296 \pi^2 \ln(2)/3
- 40624 \zeta_3/3 ,
\\
c_{4,3} &=& -65740/3 ,
\\
c_{4,4} &=&  3190 .
\eeq
The above coefficients that are associated with logarithms of $Q$
($c_{3,n}$ and $c_{4,n}$ with $n\geq 1$)
follow from the corresponding beta functions 
for $g_3$ at 2-loop \cite{twoloopQCDbeta}
and 3-loop \cite{threeloopQCDbeta} order 
and for
$y_t$ at 2-loop \cite{Fischler:1982du}, 3-loop \cite{Tarasovpreprint,Larin:1993tq} and 
4-loop \cite{Chetyrkin:1997dh,Vermaseren:1997fq} order.
Note that the 4-loop 
non-logarithmic contribution of eq.~(\ref{eq:cfourzero}) 
is only known numerically at present, with an uncertainty
from numerical integration
\cite{Marquard:2015qpa,Kataev:2015gvt}. 
The results given above are equivalent to those found in 
refs.~\cite{Tarrach:1980up,Gray:1990yh,Melnikov:2000qh,Marquard:2015qpa}, but 
are cosmetically different because the expansion in the present paper is
for the pole 
squared mass written in terms of the tree-level squared mass. The actual 
difference is of 5-loop order.

The 1-loop non-QCD contribution to the top-quark pole-squared mass can be obtained straightforwardly as:
\beq
\delta^{(1)}_{\rm non-QCD} &=& Q_t^2 e^2 t [8 - 6 \lnbar(t)] + 
\frac{g^2}{4} \Bigl \{
[2 W - b - t - (b-t)^2/W] B(b,W) 
\nonumber \\ && 
+  (2W -t + b) A(W)/W + 2 t - 2A(b) \Bigr \}
+ \bigl [\left (I_3^t \right )^2 (g^2 + g^{\prime 2})
\nonumber \\ && 
- 2 I_3^t Q_t g^{\prime 2} 
+ 2 Q_t^2 g^{\prime 4}/(g^2 + g^{\prime 2}) \bigr ]
\left [ (Z-t) B(t,Z) + A(Z) - A(t) + t \right ]
\nonumber \\ && 
+ 2 Q_t g^{\prime 2} \left [-I_3^t +  Q_t g^{\prime 2}/(g^2 + g^{\prime 2}) 
\right ]
t [3 B(t,Z) - 2]
+ 
\nonumber \\ && 
+
y_t^2 [(h-4t) B(h,t) + A(h) - 2 A(t) - A(b)]/2
- y_b^2 A(b)/2 ,
\label{eq:deltaonenonQCD}
\eeq
where $e = g g'/\sqrt{g^2 + g^{\prime 2}}$ and
\beq
Q_t = 2/3,\qquad\quad I_3^t = 1/2
\eeq
are the electric charge and third component of weak isospin for the top quark.
In eq.~(\ref{eq:deltaonenonQCD}), I have kept the contributions 
from a non-zero bottom-quark Yukawa coupling, 
but it is completely negligible in practice. This is partly 
because the leading dependence on $b$
of eq.~(\ref{eq:deltaonenonQCD}) is of order $b/t$, not $\sqrt{b/t}$ 
as in the 2-loop QCD contribution of eq.~(\ref{eq:deltatwoQCD}).

The 2-loop mixed and non-QCD contributions, $\delta^{(2)}_{\rm mixed}$ and
$\delta^{(2)}_{\rm non-QCD}$ in eq.~(\ref{eq:tpolet}), both have the form:
\beq
\sum_j c_j^{(2)} I_j^{(2)} + \sum_{j \leq k} c^{(1,1)}_{j,k} I_j^{(1)} I_k^{(1)}
+ \sum_j c_j^{(1)} I_j^{(1)} + c^{(0)} ,
\label{eq:coformc}
\eeq
where $I^{(1)}$ and $I^{(2)}$ are lists of 1-loop and 2-loop basis 
integrals defined in the conventions of \cite{Martin:2003qz,TSIL}, and 
the coefficients $c_j^{(2)}$ and $c^{(1,1)}_{j,k}$ and $c_j^{(1)}$ and 
$c^{(0)}$ consist of rational functions of the tree-level squared 
masses $t,Z,W,h$, multiplied by a global factor of $1/v^2$ for 
$\delta^{(2)}_{\rm mixed}$ and $1/v^4$ for
$\delta^{(2)}_{\rm non-QCD}$.
These coefficients are the main new results of this paper. 
However, they are complicated, and in practice will be evaluated by 
computer, so these results are relegated to ancillary electronic files. 
For both the mixed and 
non-QCD 2-loop contributions, I set the lighter quark and 
lepton Yukawa couplings to 0, 
so the list of necessary 1-loop basis integrals is:
\beq
I^{(1)} &=& 
\{A(h),\>  A(t),\>  A(W),\>  A(Z),\>  B(h, t),\>  B(t, Z) ,\>  B(0,W)\},
\label{eq:I1t}
\eeq
The list of 2-loop integrals needed for the mixed contributions is
\beq
I^{(2)}_{\rm mixed} &=&  \{\zeta_2,\> I(0,t,W),\>  I(h,t,t),\> 
I(t, t, Z),\>  M(0, 0, t, W, 0),\>  
M(0, t, t, 0, t),\>  
\nonumber \\ &&
M(0, t, t, h, t),\>  
M(0, t, t, Z, t),\>  T(h, 0, t),\> T(W, 0, 0),\>  T(Z, 0, t),\>  
\Tbar(0, 0, W),\>  
\nonumber \\ && 
\Tbar(0, h, t),\>  \Tbar(0, t, Z),\>  
U(0, W, 0, t),\>  U(t, h, t, t),\>  U(t, Z, t, t)\} 
,
\label{eq:I2mixed}
\eeq
while the 2-loop basis for the non-QCD case contains 49 additional integrals:
\beq
I^{(2)}_{\rm non-QCD} &=&  
I^{(2)}_{\rm mixed}
\cup \{
I(0, h, W),\> I(0, h, Z),\> I(0, W, Z),\> I(h, h, h),\> 
I(h, W, W),\> 
\nonumber \\ && 
I(h, Z, Z),\> I(W, W, Z),\> M(0, 0, W, W, 0),\> M(0, 0, W, W, Z),\> 
\nonumber \\ && 
M(0, t, W, 0, W),\> M(0, t, W, h, W),\> M(0, t, W, Z, W),\> M(0, Z, W, t, 0),\> 
\nonumber \\ && 
M(h, h, t, t, h),\> M(h, t, t, h, t),\> M(h, t, t, Z, t),\> M(h, Z, t, t, Z),\> 
\nonumber \\ && 
M(t, t, Z, Z, h),\> M(t, Z, Z, t, t),\> S(0, 0, 0),\> S(0, h, W),\> T(h, 0, W),\> 
\nonumber \\ && 
T(h, h, t),\> T(h, t, Z),\> T(t, h, Z),\> T(W, 0, h),\> T(W, 0, Z),\> T(W, t, W),\> 
 \nonumber \\ && 
T(Z, 0, W),\> T(Z, h, t),\> T(Z, t, Z),\> U(0, W, 0, 0),\> U(0, W, h, W),\> 
\nonumber \\ && 
U(0, W, W, Z),\> U(h, t, 0, W),\> U(h, t, h, t),\> U(h, t, t, Z),\> U(t, 0, W, W),\> 
\nonumber \\ && 
U(t, h, h, h),\> U(t, h, W, W),\> U(t, h, Z, Z),\> U(t, Z, 0, 0),\> U(t, Z, h, Z),\> 
\nonumber \\ && 
U(t, Z, W, W),\> U(W, 0, 0, h),\> U(W, 0, 0, Z),\> U(Z, t, 0, W),\> 
\nonumber \\ && 
U(Z, t, h, t),\> U(Z, t, t, Z)
\} .
\label{eq:I2nonQCD}
\eeq
The expressions for $\delta^{(2)}_{\rm mixed}$ and for 
$\delta^{(2)}_{\rm non-QCD}$ are provided in ancillary files called 
{\tt delta2mixed\_secII.txt} and {\tt delta2nonQCD\_secII.txt}, 
respectively. These files 
are available with the arXiv submission for this paper. It should be 
noted that the presentation of these 
results is not unique, because of the existence of identities 
that hold between 
different basis integrals when the squared mass arguments are not 
generic. The relevant identities are listed in the Appendix.

%%%%%%%%%%%%%%%%%%%%%%%%%%%%%%%%%%%%%%%%%%%%%%%%%%%%%%%%%%%%%%%%
\section{Re-expansion of the pole squared mass\label{sec:expandaroundpole}}
\setcounter{equation}{0}
\setcounter{figure}{0}
\setcounter{table}{0}
\setcounter{footnote}{1}

The results of the previous section can be rewritten
by self-consistently re-expanding the loop integrals that depend on $t$, 
writing them instead in terms of the real part of the pole squared mass,
\beq
T &\equiv& {\rm Re}[s_{\rm pole}].
\eeq
The resulting expression is written as
\beq
s_{\rm pole} &=& t \,+\, 
\frac{1}{16\pi^2} \bigl [g_3^2 \Delta^{(1)}_{\rm QCD} + \Delta^{(1)}_{\rm non-QCD} \bigr ]
\,+\, 
\frac{1}{(16\pi^2)^2} \left [g_3^4 \Delta^{(2)}_{\rm QCD} 
+ g_3^2 \Delta^{(2)}_{\rm mixed} + \Delta^{(2)}_{\rm non-QCD}
\right ]
\nonumber \\ &&
\,+\, 
\frac{1}{(16\pi^2)^3} g_3^6 \Delta^{(3)}_{\rm QCD} 
\,+\, 
\frac{1}{(16\pi^2)^4} g_3^8 \Delta^{(4)}_{\rm QCD} + \ldots 
\label{eq:tpoleaboutpole}
\eeq
and differs from the results of the preceding section by amounts of higher order, namely
5-loop order in the pure QCD part, and 3-loop order in the other parts. 
The pure QCD contributions are easily obtained from the results of the 
preceding section, or directly 
from refs.~\cite{Tarrach:1980up,Gray:1990yh,Melnikov:2000qh,Marquard:2015qpa}:
\beq
\Delta^{(1)}_{\rm QCD} &=& 
T [32/3 - 8 \lnbar(T)] 
,
\label{eq:Delta1QCD}
\\
\Delta^{(2)}_{\rm QCD} &=& T \Bigl [
2053/9 + 16 \pi^2/9 + 32 \pi^2 \ln(2)/9 - 16 \zeta_3/3 - (292/3) \lnbar(T) -4 \lnbar^2(T)
\nonumber \\ &&
+ \frac{16 \pi^2}{3} \sum_{q=b,c,s\ldots} M_q/ M_t
+ {\cal O}(M_q^2/M_t^2) \Bigr ]
,
\label{eq:Delta2QCD}
\\
\Delta^{(3)}_{\rm QCD} &=& T \Bigl [a_{3,0} + a_{3,1} \lnbar(T)
 + a_{3,2} \lnbar^2(T) + a_{3,3} \lnbar^3(T) \Bigr ],
\label{eq:Delta3QCD}
\\
\Delta^{(4)}_{\rm QCD} &=& T \Bigl [a_{4,0} + a_{4,1} \lnbar(T)
 + a_{4,2} \lnbar^2(T) + a_{4,3} \lnbar^3(T) + a_{4,4} \lnbar^4(T) \Bigr ],
\label{eq:Delta4QCD}
\eeq
where the 3-loop coefficients are:
\beq
a_{3,0} &=& 420365/81 + 1560884 \pi^2/1215 + 700 \pi^4/81
- 46144 \pi^2 \ln(2)/81 
\nonumber \\ &&
- 512 \pi^2 \ln^2(2)/27 - 320 \ln^4(2)/9 - (2560/3) {\rm Li}_4(1/2) 
- 12688 \zeta_3/9
\nonumber \\ &&
-11512 \pi^2 \zeta_3/27 + 31600 \zeta_5/27 
,
\\ 
a_{3,1} &=& -5648/3 - 32 \pi^2/3 - 64 \pi^2 \ln(2)/3 + 672 \zeta_3
,
\\ 
a_{3,2} &=& -36
,
\\ 
a_{3,3} &=& 8
,
\eeq
and the 4-loop coefficients are:
\beq 
a_{4,0} &=& (3.64 \pm 0.11) \times 10^5
,
\\
a_{4,1} &=& -4581172/81 - 20170532 \pi^2/1215 - 15148 \pi^4/81 + 
616000 \pi^2 \ln(2)/81 
\nonumber \\ &&
+ 6656 \pi^2 \ln^2(2)/27 + 4160 \ln^4(2)/9  + (33280/3) {\rm Li}_4(1/2) 
\nonumber \\ &&
+ 1061552 \zeta_3/27 + 149656 \pi^2 \zeta_3/27
- 598000 \zeta_5/27
,
\\
a_{4,2} &=& 11482/3 + 208 \pi^2/3 - 960 \pi^2 \ln(2) + 
            3296 \pi^2 \ln(2)/3 - 1808 \zeta_3
,
\\
a_{4,3} &=& 244/3,
\\
a_{4,4} &=& -26 .
\label{eq:Delta4QCDa44}
\eeq

The 1-loop non-QCD part has the same form as 
eq.~(\ref{eq:deltaonenonQCD}) with the replacement
$t \rightarrow T$:
\beq
\Delta^{(1)}_{\rm non-QCD} &=& 
Q_t^2 e^2 T [8 - 6 \lnbar(T)] + 
\frac{g^2}{4} \Bigl \{
[2 W - b - T - (b-T)^2/W] B(b,W) 
\nonumber \\ && 
+  (2W -T + b) A(W)/W + 2 T - 2A(b) \Bigr \}
+ \bigl [\left (I_3^t \right )^2 (g^2 + g^{\prime 2})
\nonumber \\ && 
- 2 I_3^t Q_t g^{\prime 2} 
+ 2 Q_t^2 g^{\prime 4}/(g^2 + g^{\prime 2}) \bigr ]
\left [ (Z-T) B(T,Z) + A(Z) - A(T) + T \right ]
\nonumber \\ && 
+ 2 Q_t g^{\prime 2} \left [-I_3^t +  Q_t g^{\prime 2}/(g^2 + g^{\prime 2}) 
\right ]
T [3 B(T,Z) - 2]
+ 
\nonumber \\ && 
+ y_t^2 [(h-4T) B(h,T) + A(h) - 2 A(T) - A(b)]/2 - y_b^2 A(b)/2 .
\label{eq:Delta1nonQCD}
\eeq
The 2-loop parts below absorb the 
residual terms from the expansion of $t$ about $T$,
\beq
t = T - \frac{1}{16\pi^2} \left (
g_3^2 \Delta^{(1)}_{\rm QCD} + {\rm Re}[
\Delta^{(1)}_{\rm non-QCD}] \right ) + \ldots.
\label{eq:tT}
\eeq 
Note that in eq.~(\ref{eq:Delta1nonQCD}) 
I have chosen to keep the vertex coupling $y_t$ as it is;
only the 1-loop $t$'s that come from propagators of loop integrals
are re-expanded in terms of $T$. This choice affects the residual terms 
that are absorbed into the 2-loop parts, and is somewhat arbitrary,
but is motivated by the idea 
that resumming the internal top-quark propagators in the diagram should 
result in poles close to the on-shell mass, but there is no 
reason why the vertex $y_t$'s should resum in the same way. 

The resulting 2-loop mixed and non-QCD contributions,
$\Delta^{(2)}_{\rm mixed}$ and
$\Delta^{(2)}_{\rm non-QCD}$, have the same form as 
eq.~(\ref{eq:coformc})
in the previous section,
but now the lists of necessary basis integrals 
$I^{(1)}$ and $I^{(2)}_{\rm mixed}$ and $I^{(2)}_{\rm non-QCD}$ 
are obtained from those given in the previous section by replacing $t$ by 
$T$ everywhere, including as the implicit external momentum squared  
argument of the loop integral functions.\footnote{However, note that in 
the 2-loop parts $\Delta^{(2)}_{\rm mixed}$ and $\Delta^{(2)}_{\rm 
non-QCD}$, one could justify using $T$ and $t$ interchangeably, because 
the difference is of higher order and thus formally comparable to other 
3-loop non-pure-QCD terms that remain uncalculated at this time.
Here I choose to use $T$, in solidarity with the 1-loop terms 
of eq.~(\ref{eq:Delta1nonQCD}) and the
pure QCD contributions of 
eqs.~(\ref{eq:Delta1QCD})-(\ref{eq:Delta4QCD}). This has the practical 
benefit that if $T$ is given as an input, $t$ 
can be extracted without having to re-compute the 2-loop integrals 
in iteration.\label{foot:tT}} 
In addition, the list $I^{(1)}$ used in the 2-loop parts must be 
augmented to include the real part of $B(0,W)$:
\beq
I^{(1)} &=& 
\{A(h),\>  A(T),\>  A(W),\>  A(Z),\>  
B(h, T),\>  B(T, Z) ,\>  B(0,W),\> {\rm Re}[B(0,W)]\}.
\phantom{xxxx}
\label{eq:I1T}
\eeq
The reason for this addition is that the re-expansion of 
$t$ in terms of $T$, eq.~(\ref{eq:tT}), involves the real part of
$\Delta^{(1)}_{\rm non-QCD}$. The only complex part of
$\Delta^{(1)}_{\rm non-QCD}$ is proportional to
the basis integral $B(0,W)$,
which has an imaginary part corresponding 
to the 2-body decay $t \rightarrow b W$.
The new coefficients $c_j^{(2)}$ and $c^{(1,1)}_{j,k}$ and $c_j^{(1)}$ and 
$c^{(0)}$ are now rational functions of $T,W,Z,h$, multiplied
by a factor of $1/v^2$ for $\Delta^{(2)}_{\rm mixed}$
and $1/v^4$ for  $\Delta^{(2)}_{\rm non-QCD}$.
These results are given in ancillary electronic files, 
{\tt Delta2mixed\_secIII.txt} and {\tt Delta2nonQCD\_secIII.txt}, included with the arXiv
submission for this paper. 
As in the previous section, the presentation of these results is not unique 
because of the basis loop integral identities given in the Appendix.

%%%%%%%%%%%%%%%%%%%%%%%%%%%%%%%%%%%%%%%%%%%%%%%%%%%%%%%%%%%%%%%%
\section{Numerical results\label{sec:num}}
\setcounter{equation}{0}
\setcounter{figure}{0}
\setcounter{table}{0}
\setcounter{footnote}{1}

For the purposes of a numerical illustration of the results obtained above,
consider a set of Standard Model 
benchmark $\MSbar$ parameters
\beq
y_t(Q_0) &=& 0.93690,
  \label{eq:inputyt}
\\
g_3(Q_0) &=& 1.1666,
  \label{eq:inputg3}
\\
g(Q_0) &=& 0.647550,
  \label{eq:inputg}
\\
g'(Q_0) &=& 0.358521,
  \label{eq:inputgp}
\\
\lambda(Q_0) &=& 0.12597,
  \label{eq:inputlambda}  
\\
v(Q_0) &=& \mbox{246.647 GeV},
  \label{eq:inputvev}
\eeq
defined at the $\MSbar$ input renormalization scale 
\beq
Q_0 = 173.34\> {\rm GeV}
.
\label{eq:inputQ}
\eeq
These parameters are the same as used in ref.~\cite{Martin:2015rea}. As mentioned there,
the real parts of the pole masses of the Higgs, $W$, and $Z$ bosons, as calculated in 
refs.~\cite{Martin:2014cxa},
\cite{Martin:2015lxa}, and \cite{Martin:2015rea} respectively, are:
\beq
M_h &=& 125.09 \> \rm GeV,
\\
M_W &=& 80.329 \> \rm GeV,
\\
M_Z &=& 91.154 \> \rm GeV,
\eeq
with the latter two corresponding to the experimental Breit-Wigner lineshape masses
\beq
M_W^{\rm exp} &=& 80.356 \> \rm GeV,
\\
M_Z^{\rm exp} &=& 91.188 \> \rm GeV.
\eeq
Also, although it will play no direct role in the following,
I note for completeness that the running 
$\MSbar$ Higgs squared mass parameter is 
$m^2(Q_0) = -(92.890 \> {\rm GeV})^2,$
found by using the full 2-loop effective potential 
\cite{FJJ} 
with the leading QCD and 
top-Yukawa corrections 
\cite{Martin:2013gka}
and Goldstone boson resummation \cite{Martin:2014bca,Elias-Miro:2014pca}
(see also \cite{Pilaftsis:2015bbs}), 
while one finds that 
the value obtained by including the 4-loop pure QCD corrections 
to the effective potential \cite{Martin:2015eia} is only slightly different:
$m^2(Q_0) = -(92.926 \> {\rm GeV})^2.$
For simplicity, I set $y_b = 0$, because it has a very small effect, as noted above.

\begin{figure}[!t]
\includegraphics[width=0.6\linewidth,angle=0]{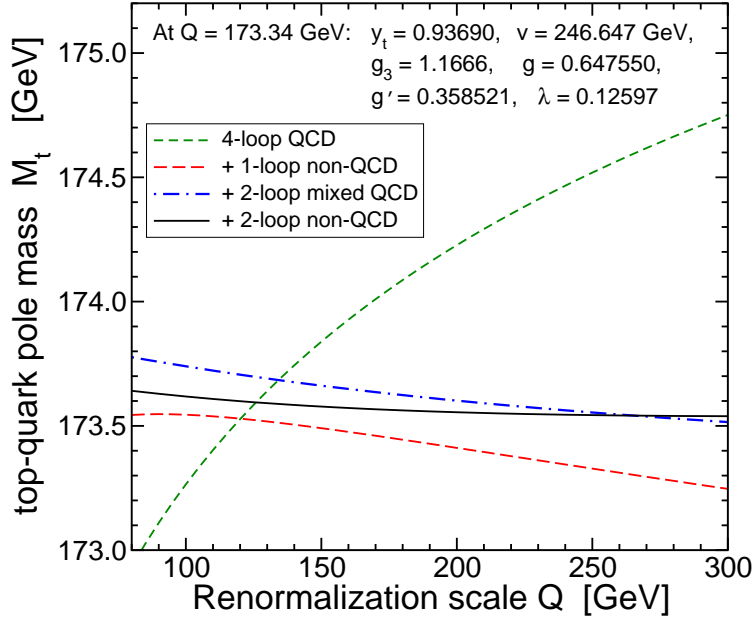}
\begin{minipage}[]{0.95\linewidth}
\caption{The real part $M_t$ of the top-quark pole mass
as a function of the $\MSbar$ renormalization scale $Q$ at which
it is computed, in successive approximations from 
section \ref{sec:expandaroundpole}. 
The short-dashed (green) line is the result found in pure QCD at 4-loop order
from using eqs.~(\ref{eq:Delta1QCD})-(\ref{eq:Delta4QCDa44}) 
in eq.~(\ref{eq:tpoleaboutpole}).
The long-dashed (red) line includes also the 1-loop non-QCD contributions
$\Delta^{(1)}_{\rm non-QCD}$ from eq.~(\ref{eq:Delta1nonQCD}).
The dot-dashed (blue) line adds in the 2-loop mixed QCD contributions
$\Delta^{(2)}_{\rm mixed}$ found 
in the ancillary file {\tt Delta2mixed\_secIII.txt}.
The solid (black) line adds in the 2-loop non-QCD contributions
$\Delta^{(2)}_{\rm non-QCD}$ found 
in the ancillary file {\tt Delta2nonQCD\_secIII.txt}.
The input parameters $y_t, v, g_3, g, g', \lambda$ at 
$Q$ are obtained by 3-loop renormalization group running
starting from eqs.~(\ref{eq:inputyt})-(\ref{eq:inputQ}).
\label{fig:MtopQ}
}
\end{minipage}
\end{figure}

Using these input parameters, the computed top-quark pole mass $M_t$ is 
shown as a function of the choice of $Q$ in figure \ref{fig:MtopQ}, in 
various approximations.
The figure was made by first using the 3-loop 
Standard Model renormalization group equations, found in 
refs.~\cite{MVI,MVII,Jack:1984vj,MVIII,Chetyrkin:2013wya,Bednyakov:2013eba} 
and as implemented in the program {\tt SMH} \cite{Martin:2014cxa}, to 
run the input parameters from the input scale $Q_0$ to the scale $Q$. 
Then, the formulas of section \ref{sec:expandaroundpole} are applied to 
find $s_{\rm pole}$, using {\tt TSIL} \cite{TSIL}. 
In addition to the integrals that are 
analytically known in terms of polylogarithms,
only 12 calls of the relatively time-consuming
Runge-Kutta evaluation function {\tt TSIL\_Evaluate} 
are needed in the full 2-loop case, because multiple basis 
integrals stemming from the same master topology are evaluated 
simultaneously. The total time to compute all of the 
basis integrals is well under 1 second on modern desktop or laptop computer hardware.
For each point, a few iterations 
are required to self-consistently evaluate the complex pole mass, 
updating $T$ with each iteration. In practical applications, the process will 
be different; one might supply $T$ as an experimental input, and derive $t$ and
therefore $y_t$ from it. 
In that case, as noted in the previous footnote, iteration of the 2-loop part 
is unnecessary when using the formulas of section \ref{sec:expandaroundpole}. 
In such applications, only the 
1-loop part will require iteration, because of the explicit appearance of $y_t$ in
eq.~(\ref{eq:Delta1nonQCD}).

\newpage
 
%%%%%%%%%%%%%%%%%%%%%%%%%%%%%%%%%%%%%%%%%%%%%%%%%%%%%%%%%%%%%%
%%%%%%%%%%%%%%%%%%%%%%%%%%%%%%%%%%%%%%%%%%%%%%%%%%%%%%%%%%%%%%

The dashed line in figure 
\ref{fig:MtopQ} shows the result of the 4-loop pure QCD calculation as given 
in eqs.~(\ref{eq:Delta1QCD})-(\ref{eq:Delta4QCDa44}) above. 
The pure QCD result for $M_t$ 
is seen to still have a significant scale dependence of 
more than 1.7 GeV for 80 GeV $<Q<$ 300 GeV, due to the effects of $y_t$ 
and the electroweak couplings. This scale dependence is 
greatly reduced by including also the 1-loop non-QCD contributions 
from $\Delta^{(1)}_{\rm non-QCD}$, as shown by the dashed (red) line. 
Further including the 2-loop mixed QCD corrections $\Delta^{(2)}_{\rm 
mixed}$, as shown by the blue (dot-dashed) line, changes the result by 
less than 300 MeV for any choice of $Q$. Finally, using the full set of 
contributions in eq.~(\ref{eq:tpoleaboutpole}) by including 
$\Delta^{(2)}_{\rm non-QCD}$ as well, one obtains the solid (black) line 
with very little $Q$ dependence. Note that with the choice $Q = M_t$,
the most complete result given here for $M_t$ 
is approximately 470 MeV lower than the 4-loop pure QCD result.

\begin{figure}[]
\includegraphics[width=0.6\linewidth,angle=0]{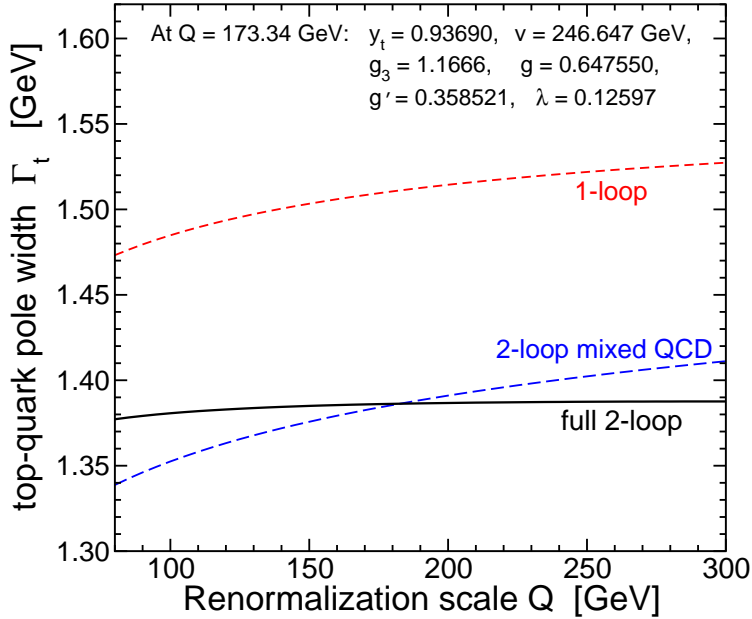}
\begin{minipage}[]{0.95\linewidth}
\caption{The pole mass width $\Gamma_t$ of the top quark,
as a function of the $\MSbar$ renormalization scale $Q$ at which
it is computed, in successive approximations from 
section \ref{sec:expandaroundpole}. 
The short-dashed (red) line is the result found from
the 1-loop non-QCD contributions
$\Delta^{(1)}_{\rm non-QCD}$ from eq.~(\ref{eq:Delta1nonQCD}),
in eq.~(\ref{eq:tpoleaboutpole}), together with the
pure QCD at 4-loop order
from eqs.~(\ref{eq:Delta1QCD})-(\ref{eq:Delta4QCDa44}).
The long-dashed (blue) line adds in the 2-loop mixed QCD contributions
$\Delta^{(2)}_{\rm mixed}$. 
%found in the ancillary file {\tt Delta2mixed\_secIII.txt}.
The solid (black) line adds in the 2-loop non-QCD contributions
$\Delta^{(2)}_{\rm non-QCD}$. 
%found in the ancillary file {\tt Delta2nonQCD\_secIII.txt}.
The input parameters $y_t, v, g_3, g, g', \lambda$ at 
$Q$ are obtained by 3-loop renormalization group running
starting from eqs.~(\ref{eq:inputyt})-(\ref{eq:inputQ}).
\label{fig:GammaTopQ}
}
\end{minipage}
\end{figure}

The complex pole mass also includes the parameter $\Gamma_t$, which 
corresponds to the total decay width of the top quark. This is shown in 
the same way as for $M_t$ in Figure \ref{fig:GammaTopQ}, again using the 
formulas in section \ref{sec:expandaroundpole}.
In this case, the width is 0 as long as only pure QCD effects are 
included, so the first approximation shown includes the 4-loop pure QCD 
together with the 1-loop non-QCD contributions to the complex pole mass, 
$\Delta^{(1)}_{\rm non-QCD}$, as the short-dashed (red) line.% 
%%%%%%%%%%%%%%%%%%%%%%%%%%%%%%%%%%%%%%%%%%%%%%%%%%%%%%%%%%%%%%
%%%%%%%%%%%%%%%%%%%%%%%%%%%%%%%%%%%%%%%%%%%%%%
The long-dashed (blue) line includes also the $\Delta^{(2)}_{\rm mixed}$ 
contribution, which lowers the prediction for the width by about 10\%, 
but without a dramatic effect on the $Q$-dependence. Finally, including the 
full 2-loop effects (on top of the pure QCD 4-loop part)  
significantly reduces the $Q$-dependence. This is shown as the solid 
(black) line in figure \ref{fig:GammaTopQ}. Of course, this result for 
$\Gamma_t$ is not as useful or complete as a fully differential 
NNLO calculation of the decay width, as described for example in 
refs.~\cite{Gao:2012ja,Brucherseifer:2013iv}
and references therein. However, it is reassuring that the result
found here is very nearly $Q$-independent.

\begin{figure}[]
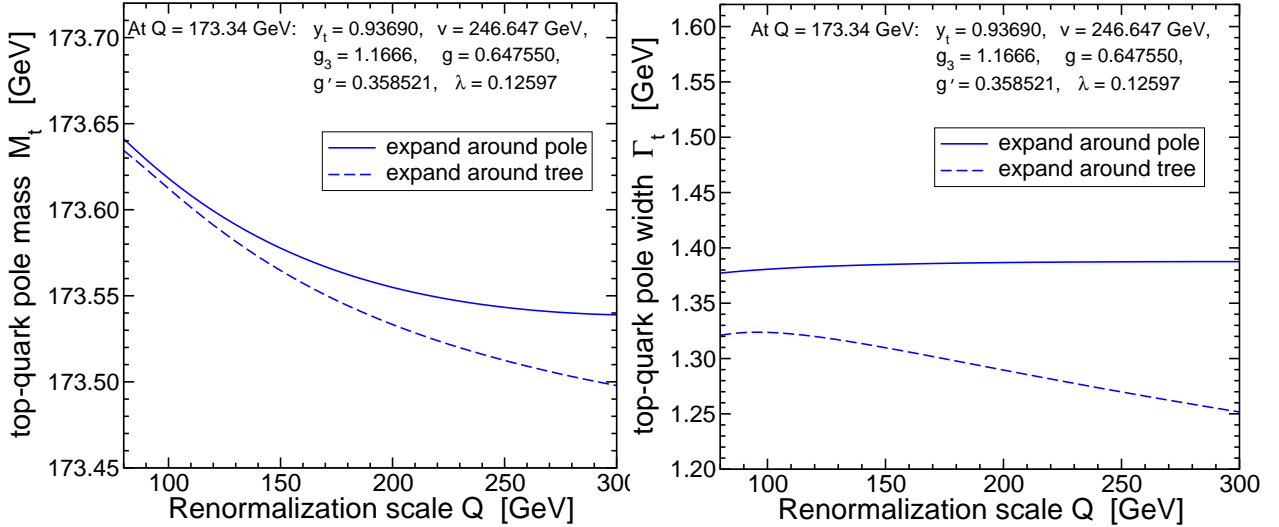

\begin{minipage}[]{0.495\linewidth}
\includegraphics[width=1.03\linewidth,angle=0]{MtopQcomp.eps}
\end{minipage}
\begin{minipage}[]{0.495\linewidth}
\begin{flushright}
\includegraphics[width=1.03\linewidth,angle=0]{GammaTopQcomp.eps}
\end{flushright}
\end{minipage}
\begin{minipage}[]{0.95\linewidth}
\caption{Comparisons of the real ($M_t$, left panel) 
and imaginary ($\Gamma_t$, right panel) parts
of the complex top-quark pole mass $M_t^2 - i \Gamma_t M_t$, 
computed using the ``expansion around tree" method
of section \ref{sec:Mtop} (dashed lines), 
and the ``expansion around pole" method of
section \ref{sec:expandaroundpole} (solid lines).
The input parameters $y_t, v, g_3, g, g', \lambda$ at the renormalization 
scale $Q$ are obtained by 3-loop renormalization group running
starting from eqs.~(\ref{eq:inputyt})-(\ref{eq:inputQ}).
The differences between the methods are formally 
of 5-loop order in pure QCD and 
3-loop order in the remaining contributions.
\label{fig:compare}
}
\end{minipage}
\end{figure}

In presenting the results above, I chose to use the expansion of section 
\ref{sec:expandaroundpole} rather than that of section \ref{sec:Mtop}. 
At least in the case of the width $\Gamma_t$, it seems clear that the 
kinematics of the decay will be best captured by using the (real part of 
the) pole squared mass $T$ rather than the running squared mass $t$ as 
the external momentum squared in the loop integrals, since in general 
$t$ is not close to the physical squared mass of the decaying top quark. 
However, the two expansions are formally equivalent within the 
approximations being used. The difference between them is a measure of 
the uncertainty introduced by the truncation of perturbation theory. 
This is illustrated in figure \ref{fig:compare}, which shows the results 
for $M_t$ and $\Gamma_t$ for the full 2-loop + pure QCD 4-loop 
expansions in terms of $t$ from section \ref{sec:Mtop} as 
dashed lines and in terms of $T$ from section \ref{sec:expandaroundpole} 
as solid lines.% 
%%%%%%%%%%%%%%%%%%%%%%%%%%%%%%%%%%%%%%%%%%%%%%%%%%%%%%%%%%%%%%
%%%%%%%%%%%%%%%%%%%%%%%%%%%%%%%%%%%%%%%%%%%%%%

The expansion in terms of $T$ in section \ref{sec:expandaroundpole} has 
a $Q$ dependence that is slightly better for $M_t$, and significantly 
better for $\Gamma_t$, than the expansion in terms of $t$ in section 
\ref{sec:Mtop}. For $\Gamma_t$, this is in accord with the expectation 
that the expansion in terms of $T$ should give a better approximation to 
the total decay width. For $M_t$, the total variation as $Q$ is varied 
from 80 GeV to 300 GeV is only about 100 MeV. As usual, the $Q$ 
dependence is only a lower bound on the theoretical error, but this 
seems to be reassuringly small compared to the experimental sources of 
error and uncertainty, at least for now. Of greater importance in the 
LHC era is the connection \cite{Mtopcollider} between the experimental 
``Monte Carlo mass" determination and the pole mass or other physical 
versions of the top-quark mass to which it can be related by other calculations.

%%%%%%%%%%%%%%%%%%%%%%%%%%%%%%%%%%%%%%%%%%%%%%%%%%%%%%%%%%%%%%%%
\section{Outlook\label{sec:outlook}}
\setcounter{equation}{0}
\setcounter{figure}{0}
\setcounter{table}{0}
\setcounter{footnote}{1}

In this paper I have presented the complex top-quark pole mass at full 2-loop order,
augmented by the known 4-loop QCD contributions, in the pure $\MSbar$ scheme. The VEV is
defined to be the minimum of the full effective potential, 
which makes it a specifically Landau gauge quantity, 
but avoids tadpole graphs. Since the VEV 
is dependent on the renormalization group scale, and therefore not a direct physical
observable anyway, 
it should not be too worrisome that it is defined to be gauge-fixing dependent.
The results found here are an alternative to the results of 
\cite{Kniehl:2015nwa}, which uses 
a tree-level definition of the running 
VEV that is independent of gauge-fixing but requires
the presence of tadpole graphs that yield powers of $1/\lambda$ in perturbative expansions.

The results obtained in this paper differ in form, even at 1-loop order, from those
found by other groups, due to the different definition of the VEV. However, one can 
check that at least the 1-loop contribution of 
eq.~(\ref{eq:deltaonenonQCD}) is consistent with, for example, eq.~(B.5) in 
ref.~\cite{Jegerlehner:2002em} or eqs.~(60) and (70) of \cite{Kniehl:2015nwa},
after taking into account eq.~(\ref{eq:translateVEVs}) of the present paper.
In ref.~\cite{Kniehl:2015nwa} it was noted that the $1/\lambda^\ell$ tadpole effects
at loop order $\ell$ can all be absorbed into a running ($Q$-dependent) 
quantity $\Delta \bar r$, defined 
in terms of the Fermi constant and $\MSbar$ quantities, including the tree-level VEV, by
\beq
G_F = \frac{1 + \Delta \bar r}{\sqrt{2}\, v_{\rm tree}^2} .
\eeq
In view of eq.~(\ref{eq:translateVEVs}) above, one can write instead, 
\beq
G_F = \frac{1 + \Delta \widetilde r}{\sqrt{2}\, v^2} ,
\eeq
where the quantity $\Delta \widetilde r$ is gauge-fixing dependent 
(because $v$ is), but free of tadpoles, and in Landau gauge is related
to $\Delta \bar r$ by the exact relation
\beq
1 + \Delta \bar r
&=& (1 + \Delta \widetilde r)\Bigl (1 + \frac{1}{\lambda v^2} 
\sum_{\ell=1}^\infty \frac{1}{(16\pi^2)^\ell}
\widehat \Delta_\ell \Bigr ).
\eeq
Some care must be taken in interpreting this, because the left side is implicitly
a function of $v_{\rm tree}$, and 
the right side a function of $v$, so that eq.~(\ref{eq:translateVEVs}) 
must be used again on the left side when making
the equivalence beyond 1-loop order. I have checked that
with this definition, $\Delta \widetilde r$ is indeed 
tadpole-free through 2-loop order, at least in the gaugeless limit for $\Delta \bar r$ 
that was presented explicitly in 
eqs.~(37)-(39) of ref.~\cite{Kniehl:2014yia}.

The 2-loop mixed and non-QCD results found in this paper 
are too complicated to show in print, 
and not amenable to unassisted human estimate anyway, so they were 
provided explicitly in electronic form in four ancillary files. 
In the near future, they will be incorporated 
into a publicly available computer program library, together with the results for the 
pole masses of the Higgs scalar and the $W$ and $Z$ bosons, as found in 
refs.~\cite{Martin:2014cxa, Martin:2015lxa, Martin:2015rea} using the same scheme
as here. (For a recent program with similar aims but based on a different organization of
perturbation theory, see
\cite{Kniehl:2016enc}.)
The QCD coupling $g_3$ is determined from other measurements. In addition, the 
QED coupling combination $gg'/\sqrt{g^2 + g^{\prime 2}}$, 
can be obtained from very low-energy experiments and renormalization group
running, and the VEV can be related to the Fermi constant $G_F$ through 
radiative corrections, in several different schemes.
In the forthcoming program, one will be able to specify either the $\MSbar$ inputs 
$y_t$, $g_3$, $g$, $g'$, $\lambda$, $v$ 
with the pole masses as outputs, or to specify the pole masses as inputs with the 
corresponding $\MSbar$ parameters as outputs, or various combinations thereof.
This program will be an extension of the Higgs mass 
program {\tt SMH} \cite{Martin:2014cxa}, and
will also include the most advanced effective potential minimization and
renormalization group running available.

%%%%%%%%%%%%%%%%%%%%%%%%%%%%%%%%%%%%%%%%%%%%%%%%%%%%%%%%%%%%%%%%
\begin{appendices}

\section{Some two-loop integral identities}\label{appendix:A}
\renewcommand{\theequation}{A.\arabic{equation}}
\renewcommand{\thefigure}{A.\arabic{figure}}
\renewcommand{\thetable}{A.\arabic{table}}
\setcounter{equation}{0}
\setcounter{figure}{0}
\setcounter{table}{0}
\setcounter{footnote}{1}

Listed below are some identities that hold between different 2-loop 
basis integrals in the notation of ref.~\cite{TSIL}, for one or more 
squared mass arguments equal to 0. The external momentum squared 
invariant is denoted $s$, and internal propagator squared masses are 
denoted $x,y,z$. In the results of section \ref{sec:Mtop}, $s$ is set 
equal to the tree-level top-quark squared mass $t$, while $s$ is set 
equal to $T = {\rm Re}[s_{\rm pole}]$ in section 
\ref{sec:expandaroundpole}.
\beq
I(0,0,x) &=& -x (1 + \pi^2/6) + A(x) - A(x)^2/2x,
\\
I(0,x,x) &=& -2 x + 2 A(x) - A(x)^2/x,
\\
S(0,x,y) &=& 
5s/8 - x - y + [
  x (x-s) T(x,0,y) 
+ y (y-s) T(y,0,x) 
\nonumber \\ && 
+ (s - x - 3 y) A(x)/2 
+ (s - 3 x - y) A(y)/2 
+ A(x) A(y) 
\nonumber \\ && 
+ (s^2 - 2 s x - 2s y + x^2 - 2 x y + y^2) B(x, y)/2
]/(s-x-y),
\\
S(0,0,x) &=& -x T(x,0,0) + (s-x) B(0,x)/2 + A(x)/2 - x + 5 s/8,
\\
U(x,0,y,z) &=&
\left[ 1/(z-y) + 1/(s-x) \right ] y T(y,x,z)
+\left[ 1/(y-z) + 1/(s-x) \right ] z T(z,x,y)
\nonumber \\ &&
+ \bigl [ 2 x T(x,y,z) + 2 S(x,y,z) - I(0,y,z) 
-A(x) -A(y) - A(z) 
\nonumber \\ &&
+x+y+z-s/4 \bigr ]/(s-x)
+ B(0,x) [A(y)-A(z)]/(z-y)
,
\\
U(x,0,y,y) &=& 
\bigl [
(s-x-4y) T(y,y,x) - 4 x T(x,y,y) - 4 S(x,y,y)
\nonumber \\ &&
+ 2 I(0,y,y) + 2 A(x) + 4 A(y)
-3x-4y+3s/2 \bigr ]/(x-s) 
\nonumber \\ &&
- [1 + A(y)/y]B(0,x) ,
\\
U(x,y,0,y) &=& 1 -T(y,0,x) + [1 - A(y)/y] B(x,y),
\\
U(x,0,0,0) &=& 1 -\overline{T}(0,0,x) + 2 B(0,x) .
\eeq
The remaining identities below hold only with $s=t$, with
$t$ being one or more of the internal propagator squared masses, as indicated.
\beq
B(0,t) &=& 1 - A(t)/t,
\\
S(0,0,t) &=& (5/8 - \pi^2/3) t + A(t)/2 - A(t)^2/2t,
\\
S(t,x,y) &=& -3 t/8 - t T(t,x,y) + [A(t) + A(x) + A(y) -x - y 
\nonumber \\ && + I(0, x, y) - x T(x, t, y) - y T(y, t, x)]/2,
\\
S(t,t,t) &=& -3 t/8 + 5 A(t)/2 - 3 A(t)^2/2 t,
\\
T(t,0,0) &=& -1 + \pi^2/3 + A(t)^2/2 t^2,
\\
T(t,0,t) &=& 
-\pi^2/18 - A(t)/t + A(t)^2/2t^2 - B(t,t)
\\
T(t,0,x) &=& 
[-(\pi^2/12) x - A(t) + A(t) A(x)/x - A(x)^2/4x]/t 
\nonumber \\ &&
- B(t, x) 
+ (x/2t -1) [T(x, 0, t) + B(t,x) + A(x)/x]
,
\\
T(t,x,x) &=& -2 - T(x, x, t) + A(t)/t - A(x)/x + A(t) A(x)/t x
\\
T(t,t,t) &=& -1 + A(t)^2/2 t^2,
\\
\overline T (0,0,t) &=& -\pi^2/3 - A(t)^2/2 t^2,
\\
U(t,0,x,y) &=& \{ 2 [t - A(t)][A(y) - A(x)]+ t [(y-x) T(t, x, y) 
\nonumber \\ && 
- 2 x T(x, t, y) + 2 y T(y, t, x)] \}/(t (x - y)),
\\
U(t,0,0,x) &=& (2 - x/2t) B(t,x) -(1 + x/2t) T(x,0,t)
+ A(x) A(t)/tx 
\nonumber \\ &&
+ A(x)^2/4tx - [1/x + 1/2t] A(x) + A(t)/t + (\pi^2/12) x/t,
\\
U(t,0,0,0) &=& 3 + \pi^2/3 - 2 A(t)/t + A(t)^2/2 t^2,
\\
U(t,0,t,t) &=& 3 - 2 \pi^2/3 - 2 A(t)/t + A(t)^2/2 t^2,
\\
U(0,t,x,y) &=& [t - x - y - I(t, x, y)]/t +
[2 A(x) A(y) - 2 y A(x) - 2 x A(y) 
\nonumber \\ &&
- x (t - x + y) T(x, 0, y) - y (t + x - y) T(y, 0, x)
\nonumber \\ &&
+ (t^2 - 2 t x  - 2 t y + x^2 - 2 x y + y^2) B(x, y) ]/(t (t - x - y)),
\\
U(0,t,t,x) &=& A(t)/t + A(x)/x 
 - A(t) A(x)/tx + [(4t - x) B(t, x) -x 
\nonumber \\ &&
 + I(0, 0, x) - 2 I(t, t, x) + (2t - x) T(x, 0, t)]/2t,
\\
U(0,t,0,x) &=& [t - x + (t-x) B(0, x) - I(0, t, x) - x T(x, 0, 0)]/t,
\\
M(0,t,t,0,t) &=& (\pi^2 \ln(2) - 3 \zeta_3/2)/t.
\eeq
Expansions in higher orders in $s-t$, which were needed in the calculations of 
this paper, are omitted for brevity but can be obtained straightforwardly from the above
by using the differential equations in $s$ listed in section IV of 
ref.~\cite{Martin:2003qz}. Similarly, expansions in small Goldstone 
boson squared masses $G$ can be obtained using the differential equations 
in section III of ref.~\cite{Martin:2003qz}.
These expansions include factors of $\lnbar(t-s)$ and $\lnbar(G)$, which 
cancel in the pole squared mass, providing useful checks.

\end{appendices}
 
\noindent {\it Acknowledgments:} 
This work was supported in part by the National
Science Foundation grant number PHY-1417028. 

%%%%%%%%%%%%%%%%%%%%%%%%%%%%%%%%%%%%%%%%%%%%%%%%%%%%%%%%%%%%%%

%%%%%%%%%%%%%%%%%%%%%%%%%%%%%%
\end{document}